\documentclass[twocolumn,prb,showpacs]{revtex4}
\usepackage{psfrag}
\usepackage{graphicx}
\usepackage{amsmath}
\usepackage{amssymb}
\def\bra#1{\mathinner{\langle{#1}|}}
\def\ket#1{\mathinner{|{#1}\rangle}}
\def\braket#1{\mathinner{\langle{#1}\rangle}}

\begin{document}
\title{A multideterminant assessment of mean field methods for the description of electron transfer in the weak coupling regime}
\author{V. Geskin$^1$, R. Stadler$^{2}$ and J. Cornil$^1$}
\affiliation{$^{1}$Laboratory for Chemistry of Novel Materials,
 University of Mons, Place du Parc 20, B-7000 Mons, Belgium\\
 $^{2}$Department of Physical Chemistry, University of Vienna,
 Sensengasse 8/7, A-1090 Vienna, Austria}

\date{\today}

\begin{abstract}
Multideterminant calculations have been performed on model systems to emphasize the role of many-body effects in the general description of charge quantization experiments. We show numerically and derive analytically that a closed-shell ansatz, the usual ingredient of mean-field methods, does not properly describe the step-like electron transfer characteristic in weakly coupled systems. With the multideterminant results as a benchmark, we have evaluated the performance of common ab initio mean field techniques, such as Hartree Fock (HF) and Density Functional Theory (DFT) with local and hybrid exchange correlation functionals, with a special focus on spin-polarization effects. For HF and hybrid DFT, a qualitatively correct open-shell solution with distinct steps in the electron transfer behaviour can be obtained with a spin-unrestricted (i.e., spin-polarized) ansatz though this solution differs quantitatively from the multideterminant reference. We also discuss the relationship between the electronic eigenvalue gap and the onset of charge transfer for both HF and DFT and relate our findings to recently proposed practical schemes for calculating the addition energies in the Coulomb blockade regime for single molecule junctions from closed-shell DFT within the local density approximation.
\end{abstract}
\maketitle

\begin{section}{Introduction}\label{sec:intro}

\begin{figure*}
    \includegraphics[width=0.95\linewidth,angle=0]{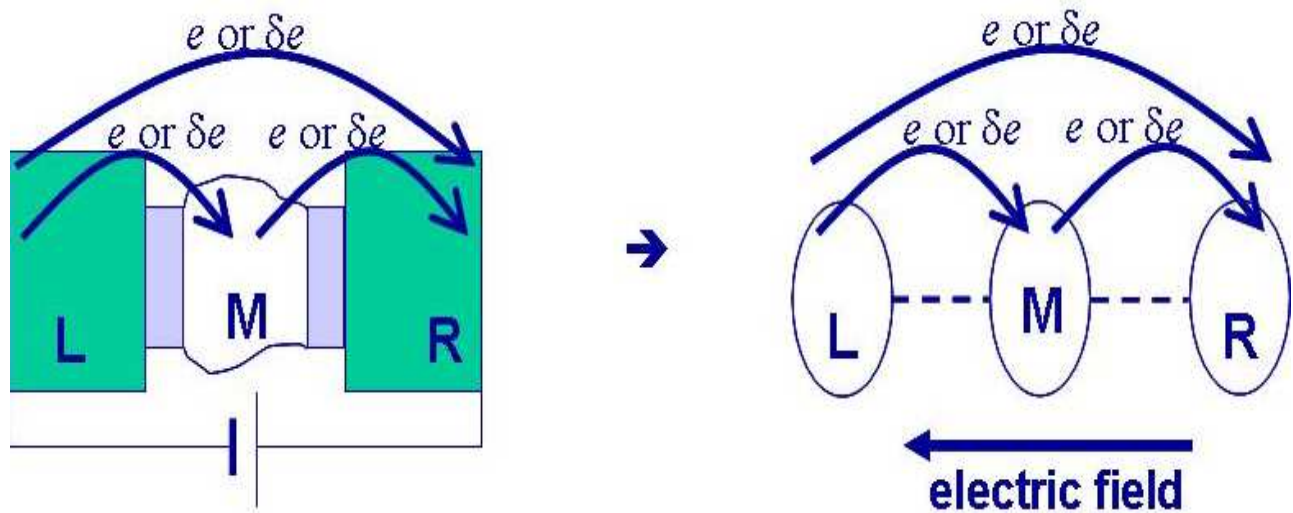}
      \caption[cap.Pt6]{\label{fig.open} (Color online) Illustration of a molecule M exchanging electrons with the left and right reservoirs (L and R) as an open system where L and R are semi-infinite contacts (left panel) and where L and R are finite (right panel).}
    \end{figure*}

Interest in electron transfer between nanoscale contacts has recently intensified, due to the advent of the technologically motivated field of molecular electronics and recent progress in experimental techniques for manipulating and contacting individual molecules~\cite{molelect,ratner}. Electron transport can operate between two limiting regimes, namely, coherent transport (CT) and Coulomb blockade (CB). While the conductivity is non-vanishing in the CT regime, CB behavior manifests itself through clear frontiers between high and low conductivity domains~\cite{kouwen}. The latter was initially reported for metallic quantum dots, where it is dominated by a capacitive charging energy. On a smaller scale, CB dominated by intrinsic level quantization has also recently been demonstrated for single molecules between electrodes~\cite{kubatkin,nitzan}. 

First-principle non-equilibrium Green's function (NEGF) methods applied to the theoretical description of electron transport through single molecule junctions are typically implemented~\cite{atk}-~\cite{kristian} in combination with density functional theory (DFT) and a closed-shell ansatz (i.e., a single determinant with double occupation of the orbitals). This approach has been used in numerous works and proved to be very useful for characterizing the CT regime, where molecules are usually closed-shell systems and remain closed-shell in the electron transport experiments. On the other hand, the description of the CB regime faces additional difficulties, as demonstrated by Datta et al.~\cite{datta} with the help of model Hamiltonians in a tight binding framework. Within this framework it was numerically shown that closed-shell mean-field models, such as DFT and spin-restricted Hartree-Fock (RHF), fail to yield the characteristic steps in electron transfer, while a spin-polarized approach such as unrestricted Hartree-Fock (UHF) might possibly yield a reasonable approximation. This is because a complete electron transfer generates an open-shell system containing a pair of singly occupied orbitals, one on the molecule and the other on the lead. A multideterminant configuration interaction (CI or Fock space ansatz) could provide a general solution to this problem. It is, however, difficult to apply a CI treatment in a first-principle description to the whole junction, especially since the metallic leads are usually represented by their mean-field band structures. This has been achieved at large computational costs with a wavefunction-based approach~\cite{greer}, where the scattering boundary conditions have been formulated in terms of the Wigner function, making an analysis of the results based on molecular orbitals quite unfeasible. Within DFT, multideterminant schemes remain quite unorthodox~\cite{GRAF05,WU07} since they diminish both its conceptual clarity and computational simplicity, and in addition require to avoid the double counting of correlation effects, which is far from trivial. Recently, a description of the CB regime was also attempted within a NEGF approach in conjunction with spin-polarized hybrid DFT~\cite{palacios,palacios1} and Hartree Fock (HF)~\cite{wang,yang} methods for junctions containing finite clusters with partially filled degenerate orbitals. 

While it is evident within wavefunction theory that a closed-shell single-determinant ansatz is inappropriate for the description of an open-shell singlet state (only a multideterminant wavefunction can ensure spin purity), the situation is less clear-cut in DFT, which is in principle an exact ground-state theory in which a singlet state does not necessarily require a spin-polarized treatment. In other words, it is expected that with an ideal XC functional, it would be in principle possible within a single-determinant Kohn-Sham (KS) DFT framework to treat strongly (non-dynamical) correlated systems, which mandatorily requires a multideterminant treatment in the framework of wavefunction theory. There are certainly grounds for such a belief since even standard functionals have been shown to include implicitly a certain degree of non-dynamic correlation\cite{GRI97,HANDY01} via their self-interaction effects~\cite{cremer,POLO02}; this suggests that the inclusion of a physical artefact might lead to an overall improvement of the results by cancelling out the effect of other approximations. Following this line of thought, hybrid functionals, with their decreased share of DFT exchange and hence disturbed balance of error cancellation empirically found for local or semi-local XC functionals, are thus expected to perform worse with non-dynamic correlation issues~\cite{COH08}. On the other hand, special functionals have been elaborated to treat bond dissociation within a spin-polarized KS-DFT formalism~\cite{FUCHS05}. Overall, it is unclear how the approaches listed above, which are commonly used to improve DFT results, will perform in the context of electron transport in the CB regime.

In this work, our goal is to gain a deeper theoretical insight into the requirements for a correct description of the charge quantization process in the CB regime by focusing on simple model systems, which allow for a direct assessment of various common mean-field methods based on HF as well as DFT against a multideterminant reference. To do so, we focus on a molecular triad as a model for an electrode-molecule-electrode system and consider electron transfer processes triggered by an external electric field, either between distinct molecules or between two moieties connected by a bridge within a single molecular unit. Due to its simplicity, our model allows us not only to compare the results of mean-field versus CI approaches on a state-of-the-art first-principle level but also to provide an analytical explanation for their qualitative differences. One of the characteristic features of the CB regime, namely steps in the dependence of the electron transfer on the applied voltage which are defined as the threshold voltages for charge injection, is recovered in our molecular model dealing exclusively with Coulomb effects and energy level quantization in the case of very weak electronic coupling between the different parts of the system. While a somewhat stronger coupling leads to a smoothening of the step-like charge transfer curves, we numerically show from first-principle calculations that the general conclusions we derive analytically for zero orbital overlap still apply. 

The paper is organized as follows: in the next section, we introduce the general setup for our model calculations and the computational techniques employed. In Section~\ref{sec:twomol}, we derive analytically and confirm numerically why and how a single-determinant closed-shell ansatz fails for the case of two molecules in  a cofacial geometry with non-overlapping orbitals and why and how a multideterminant approach repairs the deficiencies of this ansatz. Section~\ref{sec:mean} is devoted to an in-depth analysis of the capability of standard mean-field first-principle approaches to describe the intra-molecular electron transfer process in relation to the degree of conjugation between the molecular donor and acceptor parts, with a special emphasis put on spin polarization effects. Finally, Sec.~\ref{sec:summary} contains a summary.

\end{section}

\begin{section}{Methods and system setup}\label{sec:method}

\begin{subsection}{Open systems vs. finite ensembles of molecules}

As a prerequisite, it is important to first argue why our model with three molecules should share common features with an open system as encountered in CB transport experiments. The distinction between a closed and an open system depends on how one separates a region of interest from the rest of the universe~\cite{frensley}. While closed systems obey global conservation laws for mass and energy, open systems in general do not. It is frequently necessary in the context of electronic structure calculations to partition a complex system (which might be reasonably regarded as closed) into smaller components which, viewed individually, must be regarded as open. Transport phenomena are commonly described by differential equations for finite objects, with the "openness" of the system defined by the boundary conditions applied to these equations. This can be for instance achieved by employing periodic boundary conditions, which are adapted to the requirements of linear-response theory~\cite{kubo} or by defining a system which is coupled to two or more ideal reservoirs of particles~\cite{landauer}. In the latter case, the conductance is then expressed in terms of the quantum-mechanical transmission coefficients of the system in between the reservoirs, which is the basis of the non-equilibrium Green's function formalism (NEGF)~\cite{keldysh} for DFT- based electron transport calculations~\cite{atk}.

Our definition of "openness" is more elementary in the sense that the central molecule in an ensemble of three has a variable number of electrons (mass is not conserved). By applying an external field, the electrons can be transferred to the neighboring molecules, which are thus playing the role of reservoirs (Fig.~\ref{fig.open}). This is related in spirit to the statistical interpretation of fractional occupation numbers in DFT, whose physical meaning is to describe time averages of electrons exchanged between two open systems~\cite{perdew1}. The simplicity of our approach does not allow for the definition of a conductance or current, or for the characterisation of screening effects~\cite{kubatkin,first,second}. The reasons for that are two-fold: i) The absence of periodic boundary conditions for our "electrodes" prevents the unambiguous definition of a density of states or Fermi energy, which are necessary for deriving a conductance within the Landauer-B\"{u}ttiker formalism and for a steady-state description of the current; ii) in order to capture the electrostatic phenomenon leading toscreening correctly, the metallic character of electrodes (not present in our model) would be essential.

Due to these limitations of our model, we cannot really claim that we describe a CB-type transport scenario, where the key experimental quantity would be current/voltage (I/V) curves, which we are unable to produce explicitly. There are, however, spectroscopic capacitance experiments~\cite{ashoori}, which equally demonstrate gate-induced charge-quantized electron transfer without relying on a direct evaluation of I/V measurements but instead by observing discrete peaks in the device capacitance of a very sensitive transistor. Such experiments are much closer related to our model than the direct measurement of electron transport; their relation to Coulomb blockade phenomena has been clarified by B\"{u}ttiker et al.~\cite{buettiker} A common feature of both experimental approaches is the occurrence of either distinct steps in I/V curves or equivalently sharp peaks in its first derivative which are also found in the gate-dependence of the capacitance in Ref.~\cite{ashoori}. We show in our work that a correct description of such steps can be achieved by a multideterminant approach and use this solution as a benchmark for assessing the reliability of common meanfield techniques.

\end{subsection}

\begin{subsection}{Computational approaches}

In this work, we have used several computational tools as a basis for our argumentation. The Austin model 1 (AM1)~\cite{dewar}, as implemented in AMPAC~\cite{ampac}, is a wavefunction based parameterized technique, which makes it computationally very efficient for systems for which reliable atomic parameters are available, i.e., for the atoms typically found in organic molecules but only for a rather small selection of metallic atoms. The conceptual simplicity of this method and its minimal basis set make it complementary to the analytical wavefunction based models developed in Sec.~\ref{sec:twomol}. In addition to closed-shell calculations, AM1 has also been used in the framework of a complete active space configuration interaction (CASCI) scheme, in which all excited configurations  among the specified occupied and vacant RHF MOs are included in the wavefunction expansion, and their participation is determined variationally. 

In contrast, the {\it ab initio} techniques used within the GAUSSIAN package~\cite{gaussian} do not require parameters fitted to experiments, which has the advantage that the issue of transferability of such parameters to different systems or different physical boundary conditions never arises; however, this comes at the cost of a much higher computational effort. This code allows for a direct comparison (i.e., within the same computational setup and using the same basis sets) of density functional theory-based techniques, where local/semilocal or hybrid functionals are used for the exchange correlation part of the Hamiltonian, with single and multideterminant implementations of wavefunction theory, which is the main focus in Sec.~\ref{sec:mean}. The multideterminant ab initio wavefunction calculations are performed within a complete active space self-consistent field (CASSCF) scheme, in which - as a contrast to CASCI - the linear coefficients for the expansion of both the wavefunction into Slater determinants and the molecular orbitals into atomic orbitals are optimized simultaneously. Both CASCI and CASSCF active spaces include 2 highest occupied and 2 lowest unoccupied MOs.

\end{subsection}

\begin{subsection}{Model junction with three molecules}

\begin{figure}
    \includegraphics[width=0.95\linewidth,angle=0]{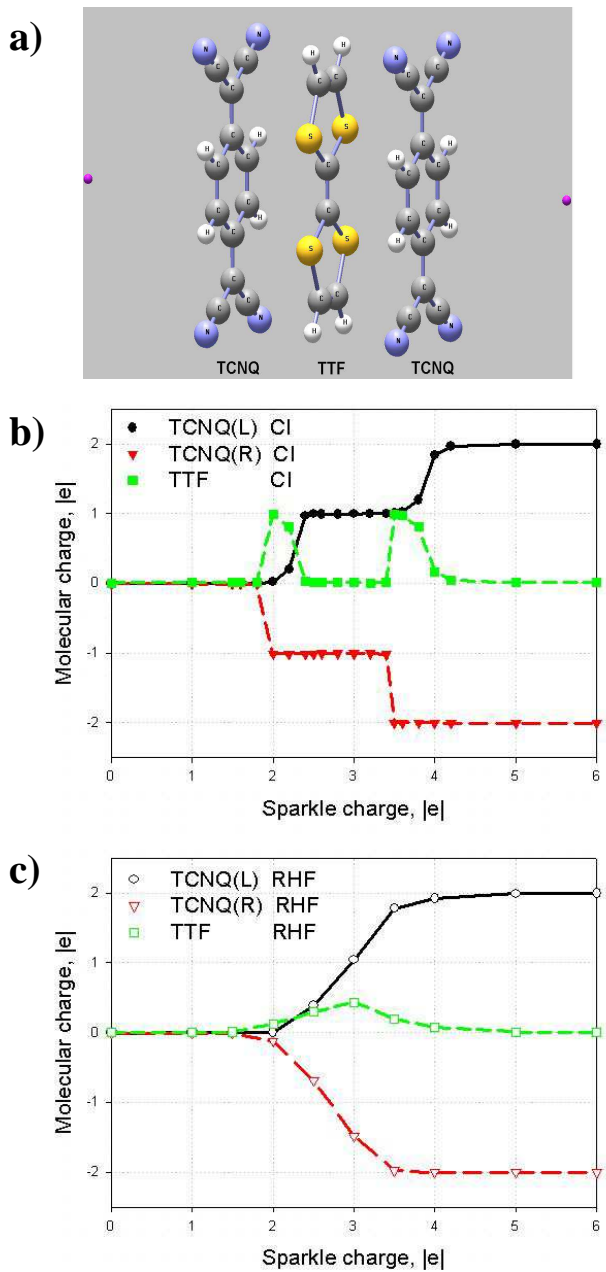}
      \caption[cap.Pt6]{\label{fig.intro} (Color online) Intermolecular charge transfer in a frozen cofacial TCNQ-TTF-TCNQ stack (a) induced by "sparkle" charges of opposite sign ($\pm$qSp) as obtained from (b) Complete Active Space Configuration Interaction (CASCI, with the active space including the highest 3 and lowest 3 MOs) and (c) RHF calculations with the AM1 Hamiltonian. TCNQ = tetracyanoquinodimethane (acceptor), TTF = tetrathiafulvalene (donor).}
\end{figure}

In Fig.~\ref{fig.intro}, we present the results of calculations performed at both the AM1/RHF and AM1/CASCI levels for the field-induced charge transfer in a TCNQ-TTF-TCNQ cofacial stack (with TTF - tetrathiafulvalene - a strong electron donor and TCNQ - tetracyanoquinodimethane - a strong electron acceptor). Although similar results can be obtained with first-principle methods, we generate them here with the parameterized Hartree-Fock AM1 Hamiltonian~\cite{ampac}, since the use of a minimal basis set facilitates our analysis in the following section. For all AM1 calculations, the external electric field is created by so-called sparkle charges~\cite{ampac} that are shown to be equivalent to a homogeneous field in the next section. The geometry of each molecule was relaxed individually, whereas the impact of inter-molecular interactions and polarization due to sparkle charges or external field on the positions of atomic nuclei have been neglected, because the focus of our article is on electronic effects only. The molecule in the middle (M) has been chosen as a donor and the molecules on the left (L) and right (R) sides as acceptors in order to reach a scenario where a significantly larger field is needed to induce the second charge transfer (M$\rightarrow$R) compared to the first (L$\rightarrow$M). In the CI description (Fig.~\ref{fig.intro}b), a series of full-electron transfer steps is found while the process is continuous with RHF (Fig.~\ref{fig.intro}c) up to the point where two electrons have been exchanged between L and R; in the latter case, only a fractional charge is localized intermediately on M. The unphysical absence of stepwise electron transfer in RHF is due to the inherent double occupation of the MOs in this theoretical framework regardless of the strength of the external field. Fig.~\ref{fig.intro} also shows that the very onset of continuous charge transfer within RHF coincides with the field strength required to induce a full one-electron transfer within CI. We will rationalize this deep correspondence by means of the analytical model introduced in the next section. If the results in Figs.~\ref{fig.intro} b and c were taken from experimental current/voltage (I/V) curves instead of theoretical calculations on electron transfer in an ensemble of molecules, one would interpret one-electron steps (as found with CI) as an indicator for the CB regime and would associate the monotonic evolution (as obtained from RHF) with CT. We thus reach the conclusion that the indiscriminate application of the closed-shell ansatz beyond its applicability may lead to a physically wrong assignment of the transport regime, which motivates a deeper analysis of the differences between the RHF and CI results.

  \begin{figure}   
  \includegraphics[width=0.95\linewidth,angle=0]{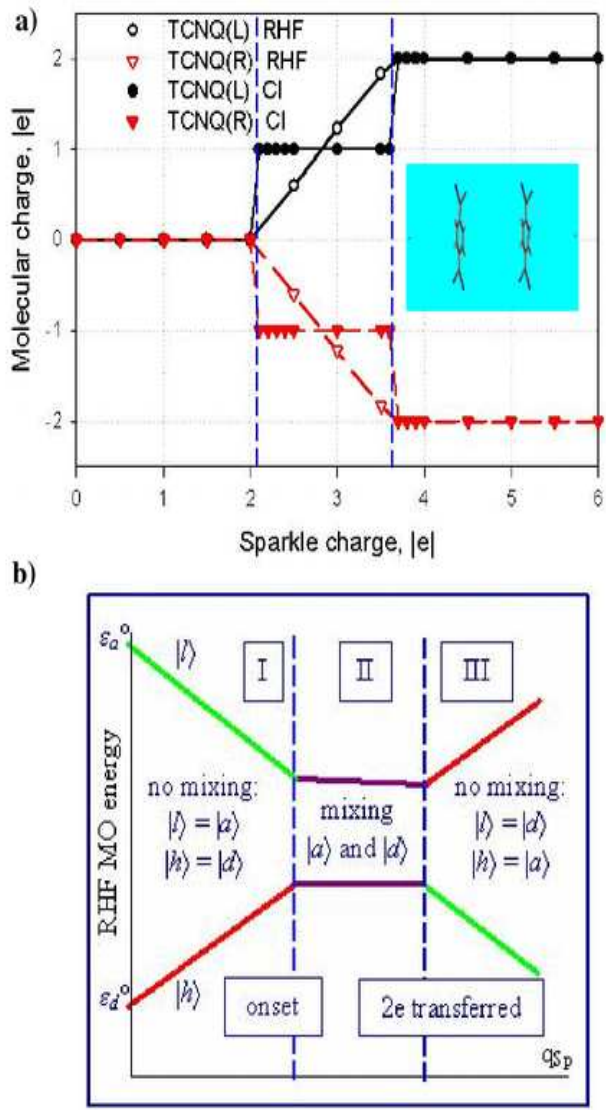}   
  \caption[cap.Pt6]{\label{fig.model}(Color online) a) Charge transfer in the TCNQ-TCNQ cofacial dyad (the molecular geometry is shown in the inset ) induced by sparkles as obtained from AM1/RHF (open symbols) and AM1/CASCI (full symbols) calculations;, b) general energy diagram obtained via a two-state model for non-overlapping molecules, see text for details.}
  \end{figure}     

\end{subsection}

\end{section}
  
\begin{section}{Electron transfer between two non-overlapping molecules in wavefunction theory}\label{sec:twomol}

\begin{subsection}{A closed-shell single-determinant approach}

When we remove the central molecule M and compute the electron transfer directly from L to R, we find the same qualitative differences between CI and RHF (Fig.~\ref{fig.model}a). We therefore concentrate hereafter on this simpler system in order to contrast the results obtained by quantum-chemical calculations within the RHF approach and within the CI formalism and to rationalize the discrepancy with the help of an analytical model.

In the field-induced charge-transfer process within the TCNQ-TCNQ dyad (Figs.~\ref{fig.model}a and b), one electron is transferred from the highest occupied MO (HOMO) of the molecule playing the role of the donor, $\ket{d}$, to the lowest unoccupied MO (LUMO) of the molecule acting as the acceptor, $\ket{a}$. In this one-electron picture, it is sufficient to consider only the HOMO $\ket{h}$ and LUMO $\ket{l}$ of the total system and the field-evolution of the weight of the $\ket{d}$ and $\ket{a}$ levels since the nature of the other orbitals of the complex does not vary with the field. The calculations show that the field-evolution of the RHF eigenenergies for $\ket{h}$ and $\ket{l}$ can be divided into three linear regions I, II, and III (see Fig.~\ref{fig.model}b). In region I, there is no charge transfer yet and the two orbitals $\ket{h}$ and $\ket{l}$ correspond to the pure HOMO and LUMO levels of the individual molecules: $\ket{h}$=$\ket{d}$, $\ket{l}$=$\ket{a}$. In region III, the two-electron charge transfer is complete, and $\ket{h}$=$\ket{a}$, $\ket{l}$=$\ket{d}$. In the intermediate region II, the gradual charge transfer found within RHF is due to a continuous mixing of the individual $\ket{d}$ and $\ket{a}$ orbitals in the frontier MOs $\ket{h}$ and $\ket{l}$ of the dyad. We discuss below the origin of this gradual orbital mixing, which in spite of the strictly zero orbital overlap between $\ket{d}$ and $\ket{a}$ ensured by the considerably large distance between the molecules, is characterized by a clear onset at the RHF level when moving from region I to II. Note that this mixing yields an energy separation between the $\ket{h}$ and $\ket{l}$ levels of about 1-2 eV throughout the entire region II of Fig.~\ref{fig.model}a. An explanation could be sought in terms of an avoided crossing scenario, as it was done for the formally similar case of molecular dissociation~\cite{daul}; however, this explanation was disproved for the latter case~\cite{cremer2} on the grounds that avoided crossings would require the size of the interaction matrix element to match the energy splitting, which in general is not the case for large intermolecular separations.

In order to rationalize these results, we consider a simple model encompassing at zero field a doubly occupied level on a donor D and a non-overlapping unoccupied level on an acceptor A, which we refer to as a Z22-model (standing for zero-overlap, 2-electron and 2-orbital). In this framework, the individual RHF eigenenergies of the isolated molecules are defined by:

\begin{eqnarray}
\label{eqn.eps}
\epsilon^0_d = & \bra{d} H_D \ket{d}  + (dd|dd) & =  \bra{d} T + V_D \ket{d} + (dd|dd) \nonumber \\
\epsilon^0_a = & \bra{a} H_A \ket{a} & =  \bra{a} T + V_A \ket{a} 
\end{eqnarray}

where T is the kinetic energy operator, V$_D$ and V$_A$ are the potential energy operators describing the interaction with the other electrons and with the nuclei on the respective molecules; $(dd|dd)$ is the two-electron Coulomb integral that appears only for the occupied orbital. When D and A form a system with zero overlap between the fragment orbitals $\ket{d}$ and $\ket{a}$, the composition of the orthonormal HOMO and LUMO orbitals $\ket{h}$ and $\ket{l}$ of the dimer are
\begin{eqnarray}
\label{eqn.hl}
\ket{h} = cos\Theta \ket{d} + sin\Theta \ket{a} \nonumber \\
\ket{l} = - sin\Theta \ket{d} + cos\Theta \ket{a} 
\end{eqnarray}

with the mixing parameter $\Theta$ characterizing the system completely. The RHF total energy defined only by the doubly occupied level $\ket{h}$ then becomes\cite{SO96}:
\begin{eqnarray}
\label{eqn.etot}
E_{RHF}  = 2 \bra{h} T + V_D + V_A + V_{bias} \ket{h} + (hh|hh) = \nonumber \\
2\epsilon^0_d - (dd|dd) + 2q\kappa_D 
+2sin^2 \Theta (\epsilon^0_a-\epsilon^0_d - (dd|aa) \nonumber \\
- q(\kappa_A + \kappa_D ))+sin^4 \Theta ((dd|dd)+(aa|aa)-2(dd|aa))   
\end{eqnarray}
The external field enters E$_{RHF}$ via V$_{bias}$ which depends on the magnitude of the sparkle charges q (see Fig.~\ref{fig.intro} for details) and on the spatial distribution of the occupied orbital via an effective coefficient $\kappa$, that is V$_{bias}$(A,D) = q$\kappa_{A,D}$. 

There is no mixing of the D and A fragment orbitals if sin$^2\Theta$=0, i.e., $\ket{h}$=$\ket{d}$ and $\ket{l}$=$\ket{a}$ (region I in Fig.~\ref{fig.model}b). Therefore, the degree of RHF mixing can be obtained by minimizing E$_{RHF}$ in Eq.~\ref{eqn.etot} with respect to sin$^2\Theta$, where
\begin{eqnarray}
\label{eqn.theta}   
sin^2\Theta = \frac{(dd|aa) - (\epsilon^0_a-\epsilon^0_d) + (\kappa_D + \kappa_A )q}{((dd|dd) - (dd|aa)) + ((aa|aa) - (dd|aa))} 
\end{eqnarray}
The condition sin$^2\Theta\geq$0, which entirely determines the onset of RHF mixing, is satisfied only when the numerator of Eq.~\ref{eqn.theta} is non-negative (the denominator is positive as one-center Coulomb integrals in general exceed two-center integrals), that is when
\begin{eqnarray}
\label{eqn.cond}   
(\epsilon^0_a - \kappa_A q) - (\epsilon^0_d + \kappa_D q) \leq (dd|aa). 
\end{eqnarray}
Eq.~\ref{eqn.cond} means that the onset of mixing (the point at the crossing of region I and region II in Fig.~\ref{fig.model}b) takes place when the energy difference between the LUMO of the acceptor and the HOMO of the donor in its linear decrease under the influence of the sparkle charges $q$, equals the Coulomb integral $(dd|aa)$. 
Within RHF the eigenenergies of the vacant and occupied MOs approximate the electron affinity EA and ionization potential IP, respectively, due to Koopmans theorem. On the other hand, the integral $(dd|aa)$ is the Coulomb interaction energy between the charge densities of the two molecules, which are well separated in space in our case. This Coulomb integral is expected to evolve as 1/d for the asymptotic limit of large intermolecular distances, which is verified in Fig.~\ref{fig.distance} showing the distance dependence of the HOMO-LUMO gap of the dyad at the onset of the charge transfer as calculated numerically with AM1. Therefore, in the long distance limit, the RHF mixing onset is also the point where the difference between EA and IP is fully compensated by the Coulomb attraction of the formed ions, thereby generating the necessary conditions for a full one-electron transfer, which occurs within the correct CI treatment at the same bias in Fig.~\ref{fig.model}a.

It is straightforward to demonstrate from the standard expression of the RHF energies of the occupied and vacant molecular orbitals~\cite{SO96}:

\begin{eqnarray}
\label{eqn.gap}
\epsilon_h = \bra{h} t + V_D + V_A + V_{bias} \ket{h} + (hh|hh) \nonumber \\
\epsilon_l = \bra{l} t + V_D + V_A + V_{bias} \ket{l} + 2 (hh|ll) + (hl|hl)
\end{eqnarray}

and making use of Eqs.~\ref{eqn.hl} and~\ref{eqn.theta} that the HOMO-LUMO gap actually remains constant and equal to $(dd|aa)$ throughout the mixing region II, though the eigenenergies $\epsilon_h$ and $\epsilon_l$ may change.

We note further that the Mulliken charge~\cite{mulliken} Q on the non-overlapping fragments D and A is proportional to the square of the coefficients in the orbital expansion of $\ket{h}$ multiplied by $\ket{h}$'s occupation (which is always 2), that is $\vert$Q$\vert$=2 sin$^2\Theta$. According to Eq.~\ref{eqn.theta}, Q is linear with the external field generated by the sparkle charges q, which explains the continuous intermolecular charge transfer obtained from closed-shell calculations (Fig.~\ref{fig.model}a).

\begin{figure}   
\includegraphics[width=0.95\linewidth,angle=0]{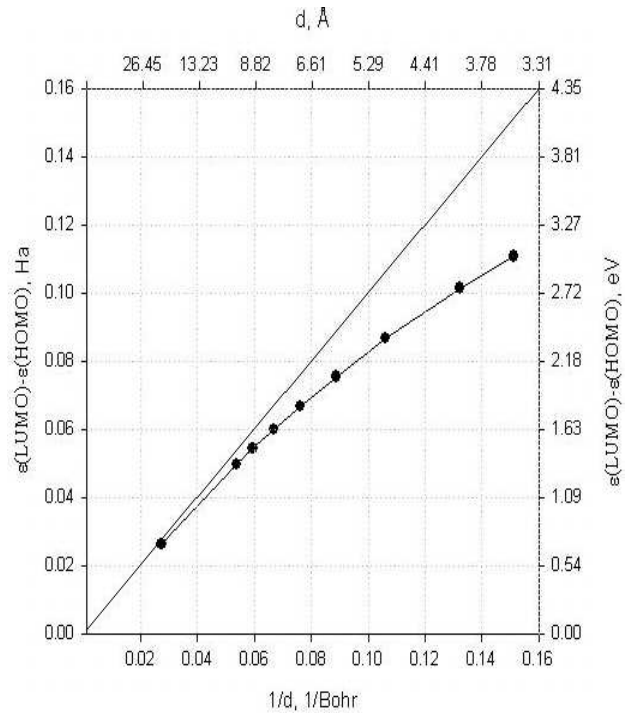}   
\caption[cap.Pt6]{\label{fig.distance}(Color online) Evolution of the threshold HOMO/LUMO gap for the onset of charge transfer in the AM1/RHF calculations (i.e., region II in Fig.~\ref{fig.model}b) in a TCNQ-TCNQ cofacial dyad as a function of the intermolecular distance d. The diagonal line shows the point-charge interaction energy, evolving as E=1/d (in atomic units e=1, 1Hartree(Ha)=1/Bohr).}
\end{figure}     

\end{subsection}

\begin{subsection}{Multideterminant approach}

The open-shell singlet wavefunction describing the final state of an electron transfer from D to A can be written as
\begin{eqnarray}
\label{eqn.psione}
\Psi(D^+A^-)=\frac{1}{\sqrt{2}}(|da|-|ad|) \nonumber \\
=\frac{1}{\sqrt{2}}(da+ad)(\uparrow\downarrow-\uparrow\downarrow)	
\end{eqnarray}
where $|da|$ is a Slater determinant, and $(da+ad)\equiv (\ket{d(1)}\ket{a(2)}+\ket{a(1)}\ket{d(2)})$ and $(\uparrow\downarrow-\downarrow\uparrow)\equiv(\ket{\uparrow(1)}\ket{\downarrow(2)}-\ket{\downarrow(1)}\ket{\uparrow(2)})$ are the shorthand notations for the spatial and spin parts of the two-electron wavefunction, respectively. In the following, we illustrate how to construct $\Psi$ from the RHF-MOs $\ket{h}$ and $\ket{l}$ introduced in the last section or in other words how to move from a single determinant (RHF) to a multideterminant (CI) description, where the Fock space is a minimal CI space encompassing four determinants generated by single and double excitations from $\ket{h}$ to $\ket{l}$. 

It is straightforward to show that, in the Fock space spanned by four Slater determinants, namely, $|hh|$, $|ll|$, $|hl|$ and $|lh|$, a multideterminant singlet wavefunction of the general form 
\begin{eqnarray}
\label{eqn.psitwo}
\Psi(D^+A^-)=x(|hl|-|lh|)+y(|hh|-|ll|) \nonumber \\ 
=[(-x sin(2\Theta)+y cos(2\Theta))(dd-aa) \nonumber \\
+(x cos(2\Theta)+y sin(2\Theta))(da+ad)](\uparrow\downarrow-\uparrow\downarrow)
\end{eqnarray}
with the MOs $\ket{h}$ and $\ket{l}$ expressed as in Eq.~\ref{eqn.hl}, corresponds to $\Psi$ in Eq.~\ref{eqn.psione}, for the whole range of the mixing parameter $\Theta$, if the conditions
\begin{eqnarray}
\label{eqn.coeff}
x = \frac{1}{\sqrt{2}} cos(2\Theta), y =\frac{1}{\sqrt{2}} sin(2\Theta)
\end{eqnarray}
are fulfilled.  We stress that $\Psi(D^+A^-)$, in order to remain constant according to Eq.~\ref{eqn.psione}, has to vary continuously when expressed in the basis of the molecular orbitals $\ket{h}$ and $\ket{l}$, which are variables themselves due to RHF mixing. For the mixing parameter $\Theta$ at the onset and in the middle of the region II of Fig.~\ref{fig.model}, we find by combining Eqs.~\ref{eqn.psitwo} and ~\ref{eqn.coeff} 
\begin{eqnarray}
\label{eqn.psithree}
\Psi(D^+A^-) & = & (1/\sqrt{2})(|hl|-|lh|) \quad \mathrm{for} \, \Theta\rightarrow0 \nonumber \\
\Psi(D^+A^-) & = & (1/\sqrt{2})(|hh|-|ll|) \quad \mathrm{for} \, \Theta=\pi/4.
\end{eqnarray}
 
\begin{figure}   
\includegraphics[width=0.95\linewidth,angle=0]{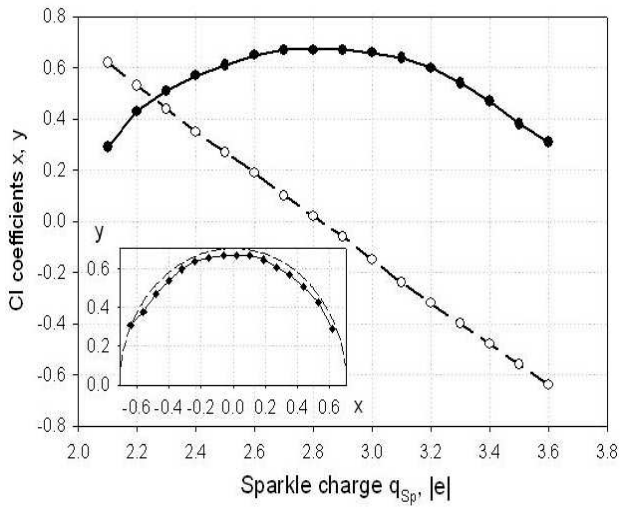}   
\caption[cap.Pt6]{\label{fig.fournew}(Color online) CI coefficients x and y appearing in Eq.~\ref{eqn.psitwo} as obtained from AM1/CASCI calculations for a TCNQ-TCNQ cofacial dyad as a function of the sparkle charge in the RHF mixing region II (see Fig.~\ref{fig.model}). Inset: y vs. x as obtained from these calculations compared to the prediction of the Z22 model.}
\end{figure}     

Whereas the RHF determinant $|hh|$ dominates the wavefunction before the RHF mixing (region I), the CI wavefunction does not contain a relevant contribution from it just at the onset (see Eq.~\ref{eqn.psithree}). This abrupt change in the nature of the electronic ground state, pointing to the inadequacy of RHF to provide a physically correct solution to the problem, is due to the crossing of $D^0A^0$ and $D^+A^-$ at the threshold value of $q$. It is mainly the total energy of $D^+A^-$ that strongly depends on $q$ due to its polarity, whereas the energy of $D^0A^0$ is unaffected by $q$ since the out-of-plane polarizability of the planar TCNQ molecule is negligible. The crossing of the ground and first excited states is not avoided in the zero wavefunction overlap limit. We will show in the next Section that the abruptness of this transition is smoothened when introducing the overlap between the donor and acceptor orbitals.

It is clear from Eq.~\ref{eqn.coeff} that the CI coefficients x and y are determined by the external field parameter q via the orbital mixing parameter $\Theta$. In order to obtain the details of this dependency, we note that $cos(2\Theta) = 1 - sin^2\Theta$ decreases linearly with the external field (according to Eq.~\ref{eqn.theta}, $sin^2\Theta$ is linear with q in the RHF orbital mixing region II). Therefore, it follows from Eq.~\ref{eqn.coeff} that the coefficient $\pm$x varies linearly with q between -1/$\sqrt{2}$ and 1/$\sqrt{2}$ and its counterpart $\pm$y is defined by $y = \sqrt{1/2 - x^2}$. In Fig.~\ref{fig.fournew} we plot the coefficients x and y as a function of the parameter q for the external field, as calculated at the corresponding AM1/CASCI level for the TCNQ cofacial dimer within the mixing region found in Fig. 3a. This data validates the functional dependencies derived from our analytical model.

\end{subsection}

\begin{subsection}{Discussion}

Our analytical model demonstrates that a closed-shell single-determinant RHF approach wrongly predicts continuous charge transfer between non-overlapping moieties, which is automatically corrected to integer electron transfers with a minimal multideterminant ansatz. While RHF finds variationally the best wavefunction in the form of a single determinant based on doubly occupied MOs, this solution is totally inappropriate for the description of the open-shell singlet state reached by single electron transfer processes. Such an open-shell singlet state can only be described by a series of determinants based on RHF MOs. The coefficients in the expansion introduced in Eq.~\ref{eqn.psitwo} add the additional variational degrees of freedom to the model, so that the resulting multideterminant ansatz properly describes the system. Our Z22 model explains: (i) the sharp onset of continuous charge transfer in RHF as a function of the external field (which is different from a conventional avoided crossing); (ii) the reason for the unphysical linearity of the field-dependence of the charges on the donor and acceptor units in RHF; (iii) the origin of the HOMO-LUMO gap in the RHF mixing region; (iv) why the RHF charge-transfer onset coincides with the full one-electron transfer in CI; and finally (v) how CI corrects for the described artefacts of RHF.

As the energy splitting between the two frontier MOs $\ket{d}$ and $\ket{a}$ gets reduced with the increase in the electric field, the system as a whole becomes strongly correlated, or in other words non-dynamic correlation becomes essential~\cite{HUR76,MOK96,HANDY01}; the latter is also frequently referred to as static, left-right or near-degenerate correlation, and is also crucial for a correct description of the transition point in molecular dissociation processes~\cite{FUL02,GOK08}. There are similar methodological problems in the case of the dissociation of ionic molecules, e.g. LiF, where deficiencies in RHF or DFT calculations lead to (partially) charged atoms at large separation, while a correct treatment does always yield neutral open-shell atoms after the crossover in the total energies of the neutral and ionic states~\cite{BAU88,HAN08}. 

Our results were derived so far for the case of zero orbital overlap between the two molecules. In the next section, we  allow for a weak overlap between the orbitals of the donor and acceptor moieties and show that this does not alter the general conclusions derived here.

\end{subsection}

\end{section}
    
\begin{section}{Mean field approximations for intramolecular electron transfer}\label{sec:mean}

\begin{subsection}{Aviram Ratner molecules}

After having compared single- versus multideterminant approaches for a simple system both at an analytical and semi-empirical level, we devote the current section to approaching the same problem with common mean-field techniques based on a first-principle description; all calculations reported hereafter were performed with the Gaussian package~\cite{gaussian}.We have investigated different model systems to vary the coupling strength between the donor and acceptor moieties determined by the degree of conjugation provided by the bridging unit and assess its influence on the nature of the electron transfer (i.e., whether it is continuous, stepwise or intermediate in its dependence on the external electric field is not straightforward to predict a priori). In particular, we consider the Aviram-Ratner molecule~\cite{AR74}, where the acceptor TCNQ and the donor TTF units are linked by a non-conjugated bridge denoted TCNQ-PHP-TTF, and its shorter analog with a conjugated bridge denoted TCNQ-PP-TTF. For these systems, the applied field is not perpendicular to the molecular planes of the D and A moieties as in the previous section but on the contrary directly along the molecular backbone so that the gradual polarization of the $\pi$ electron system is also expected to play a role in the physics of the electron-transfer process.

\begin{figure}
      \includegraphics[width=0.95\linewidth,angle=0]{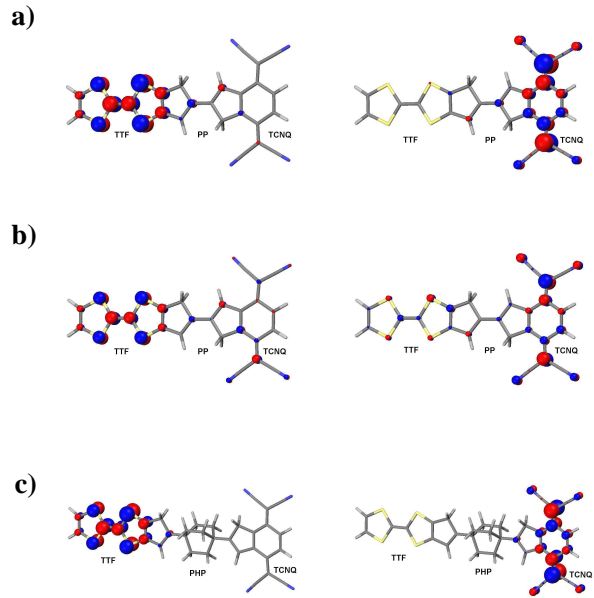}
      \caption[cap.Pt6]{\label{fig.mo} (Color online) Shape of the frontier orbitals (HOMO in the left and LUMO in the right panels, respectively) at zero field for TCNQ-PP-TTF as obtained with a) RHF; b) DFT-BLYP and c) for TCNQ-PHP-TTF with RHF using in all cases a 6-31G** basis set. The size and color of the balls are representative of the amplitude and sign of the LCAO coefficients, respectively.}
\end{figure}

In the absence of any external field, the frontier MOs are completely localized on the D and A parts for the longer (PHP) bridge (Fig.~\ref{fig.mo}). For the shorter (PP) bridge, there is some spill-over in the HOMO and LUMO orbitals with RHF and even appreciable delocalization for DFT using semi-local XC functionals (hybrid functionals yield a behaviour intermediate between semi-local DFT and RHF). As expected, we also note that the HOMO-LUMO gap is much larger in the RHF than in the DFT description providing a negligible value. Within standard DFT, we are dealing with a peculiar artefact for this system in the sense that the strongly reduced gap of the isolated TTF and TCNQ molecules makes the LUMO of the acceptor lie lower than the HOMO of the donor. This can only be remedied by hybrid functionals with a sufficient share of HF exchange (e.g. 50\% in the hybrid BHandHLYP potential)~\cite{AVI09}.

We stress that the intra-molecular electron transfer falls in the weak coupling or Coulomb blockade regime for the two molecules studied, although they differ by their internal degree of coupling provided by the PP and PHP bridges. Low thresholds and smooth charging curves are physically correct observables in the strong coupling regime of coherent electron transport since CASSCF calculations predict there a continuous partial electron transfer reflecting the change in the spatial distributions of the strongly overlapping (and hence delocalized) frontier orbitals. This scenario is, however, beyond the scope of this paper since our model derived in Sec.~\ref{sec:twomol} is based on the assumption of zero orbital overlap, which makes it unsuitable for predictions in the coherent transport regime.

\end{subsection}

\begin{subsection}{Single-determinant closed-shell methods: non-spin-polarized DFT and RHF}

\begin{figure}
      \includegraphics[width=0.95\linewidth,angle=0]{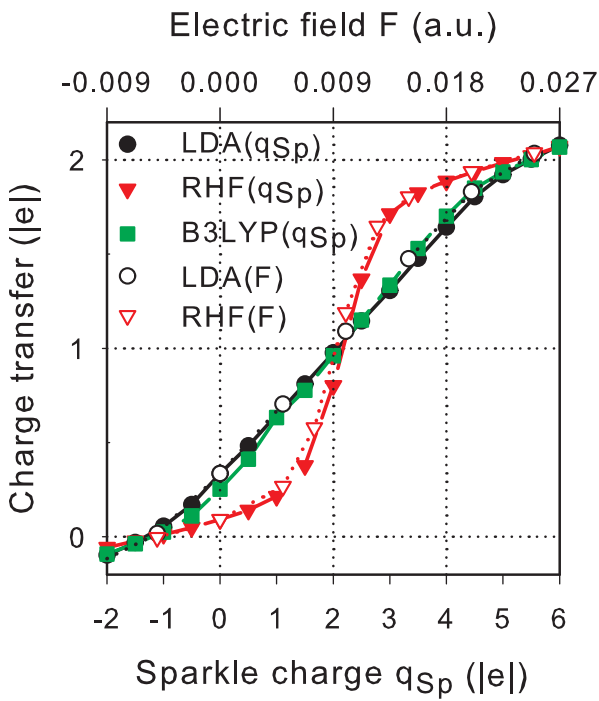}
      \caption[cap.Pt6]{\label{fig.central} (Color online) Intramolecular charge transfer between the donor and acceptor moieties in TCNQ-PP-TTF induced by sparkle charges (dashed lines) or an external electric field (dotted lines) as obtained by {\it ab initio} closed-shell pure DFT at the LDA level (circles), hybrid DFT at the B3LYP level (squares), and RHF calculations (triangles) with a 6-31G** basis set. The significant zero-field charge transfer between the moieties obtained in the LDA calculations originates from an artefact: for the isolated molecules, the LUMO of the acceptor TCNQ lies much lower than the HOMO of the donor TTF (-5.44 versus -3.90 eV) unlike RHF results (-1.97 and -6.77 eV); note that the gas-phase EA of TCNQ and IP of TTF, 2.8 annd 6.8 eV, respectively~\cite{nist}.}
\end{figure}

Fig.~\ref{fig.central} shows the charge separation between the donor and acceptor parts in the TCNQ-PP-TTF molecule as obtained from first-principle calculations within DFT at the local density approximation (LDA) level and with the hybrid exchange correlation (XC) functional B3LYP (with 20\% of HF exchange) as well as with RHF for a direct comparison with wavefunction theory. The data in Fig.~\ref{fig.central} further demonstrates the complete equivalence of the polarization induced by sparkle charges and by the application of an external electric field, thus justifying the use of sparkle charges throughout this work. We display in Figs.~\ref{fig.PP} and~\ref{fig.PHP} results calculated from wavefunction theory (Figs.~\ref{fig.PP}a,~\ref{fig.PHP}a) and DFT (Figs.~\ref{fig.PP}b,~\ref{fig.PHP}b) for a molecule introducing a strong coupling between the D and A units (Fig.~\ref{fig.PP}- TCNQ-PP-TTF) and a weaker coupling (Fig.~\ref{fig.PHP}- TCNQ-PHP-TTF), respectively. In these figures, we also contrast the closed-shell results with calculations allowing for spin polarization (see next section) at both the RHF and DFT levels and a coupling with a CI scheme in the case of RHF (Figs.~\ref{fig.PP}a,~\ref{fig.PHP}a).
The general features of the ab initio closed-shell RHF curves for intramolecular electron transfer in Figs.~\ref{fig.central},~\ref{fig.PP}a and~\ref{fig.PHP}a are reminiscent to those encountered in the model dyad with AM1 (see Fig.~\ref{fig.model}). Whereas there is some smoothening at the onsets due to orbital overlap in TCNQ-PP-TTF (Figs.~\ref{fig.central},~\ref{fig.PP}a), the absence of such overlap in TCNQ-PHP-TTF (Fig.~\ref{fig.PHP}a) makes the mixing region as sharply separated from the integer-charge regions as in the model systems. A step-like behavior is found with CI and a continuous change of the Mulliken charges with closed-shell RHF, as observed in Figs.~\ref{fig.intro} and~\ref{fig.model}.  The closed-shell DFT-LDA charging curves are fully linear with the field, which is due to the possibility of fractional charges in the framework of DFT; using the hybrid B3LYP or BHandHLYP functionals with 20\% and 50\% of HF exchange, respectively, does not modify this behavior. 

It has been recently suggested that a standard DFT framework has inherent problems for describing electron transport in both the CB and CT regimes~\cite{sanvito1}, due to the self-interaction of electrons~\cite{perdew} in a Kohn-Sham framework and the lack of a derivative discontinuity (DD)~\cite{perdew1} in the evolution of the HOMO eigenenergy as a function of its occupancy, which can be fractional. In Ref.~\cite{sanvito1}, these two issues have been portrayed as intimately linked and a SI correction scheme has been devised as a remedy. In particular, Ref.~\cite{sanvito1} shows two distinct things: i) calculations with a model discontinuous potential lead to the recovery of steps in the electron occupation in a model quantum dot in the weak coupling regime, while a continuous potential induces a continuous dependence on voltage, thus demonstrating the role of DD; ii) effective tight-binding transport calculations on molecular system contacted by two probes suggest that the corrections for SI could serve to introduce the desired discontinuities into the XC functional and thereby produce CB steps. In a more recent paper from the same group~\cite{TS08}, it is argued that many failures of standard DFT can be traced back to SI, though the authors also concede that the physical reasons for the lack of DD in common XC functionals are too complex to be explained by SI alone. It is therefore an open issue to assess how much of the problems DFT faces in describing transport in the CB regime can be attributed to SI; note that Datta~\cite{datta} has also stressed that SI is equally absent in RHF and UHF, therefore not offering an explanation why the latter performs better than the former. We showed above that the incorrect description of steps in the charge/voltage curves related to a lack of DD in Ref.~\cite{sanvito1} is actually a general feature of closed-shell methods. Since RHF is by definition completely SI free, the RHF evolution in Figs.~\ref{fig.model},~\ref{fig.central},~\ref{fig.PP}a and~\ref{fig.PHP}a exemplifies the limits for the improvements that can be achieved by SI correction techniques within non spin-polarized KS-DFT, at least with commonly used XC functionals.

\end{subsection}

\begin{subsection}{Single-determinant open-shell methods: spin-polarized DFT and UHF}\label{sec:uhf}

\begin{figure}[!htb]
      \includegraphics[width=0.95\linewidth,angle=0]{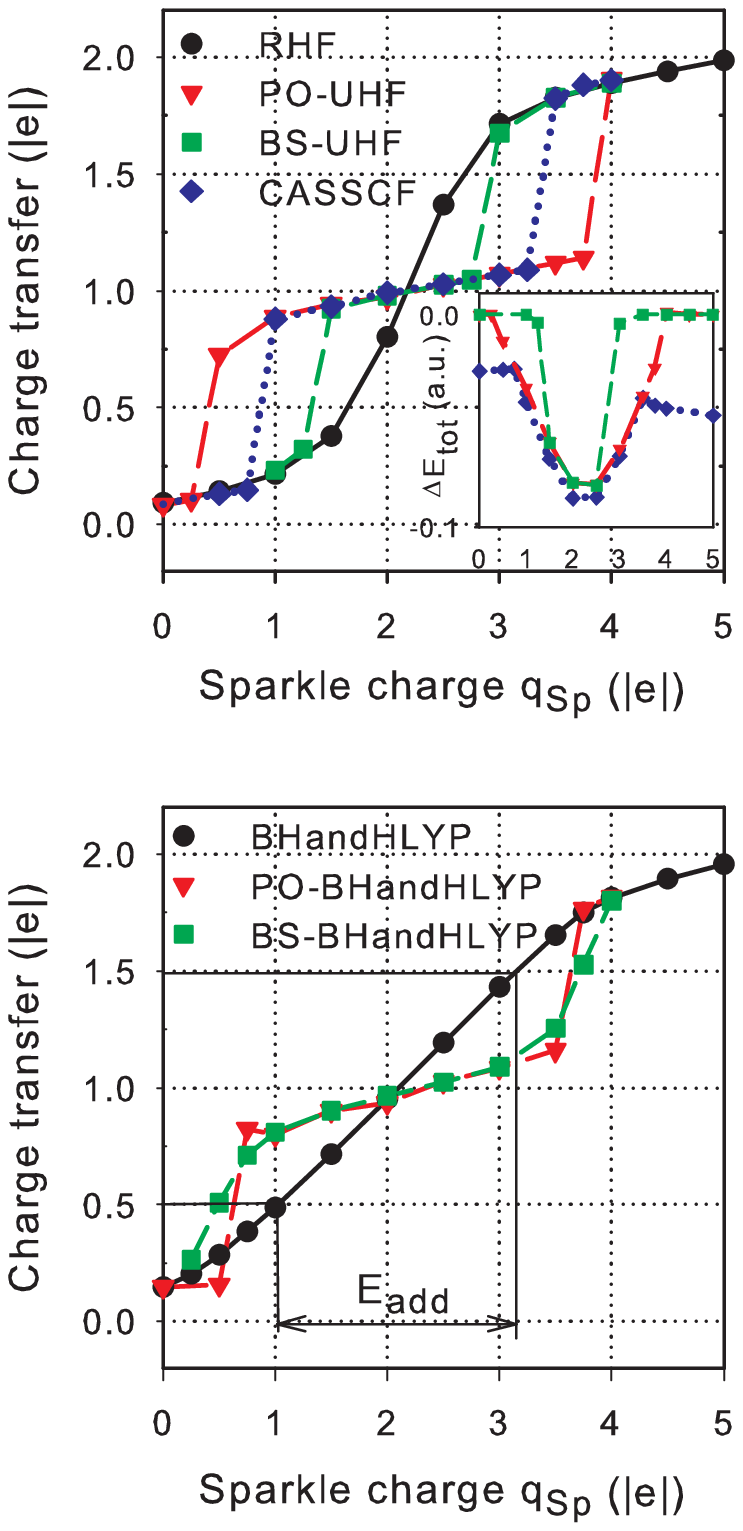}
      \caption[cap.Pt6]{\label{fig.PP} (Color online) Intramolecular charge transfer between the donor and acceptor moieties in TCNQ-PP-TTF induced by sparkle charges as obtained by: a) {\it ab initio} wavefunctions; and b) hybrid DFT with a 6-31G** basis set. In a) restricted (circles), PO-unrestricted (triangles) and BS-unrestricted (squares) HF and CASSCF (diamonds, with the HOMO and LUMO included in the active space) results are compared. The inset shows the energy gain of the three latter methods vs. RHF. In b) restricted (circles), PO- (triangles) and BS-unrestricted BHandHLYP (squares) results are compared; addition energies E$_{add}$ defined in Refs.~\cite{first,second} are also illustrated.}
\end{figure}

\begin{figure}[!htb]
      \includegraphics[width=0.95\linewidth,angle=0]{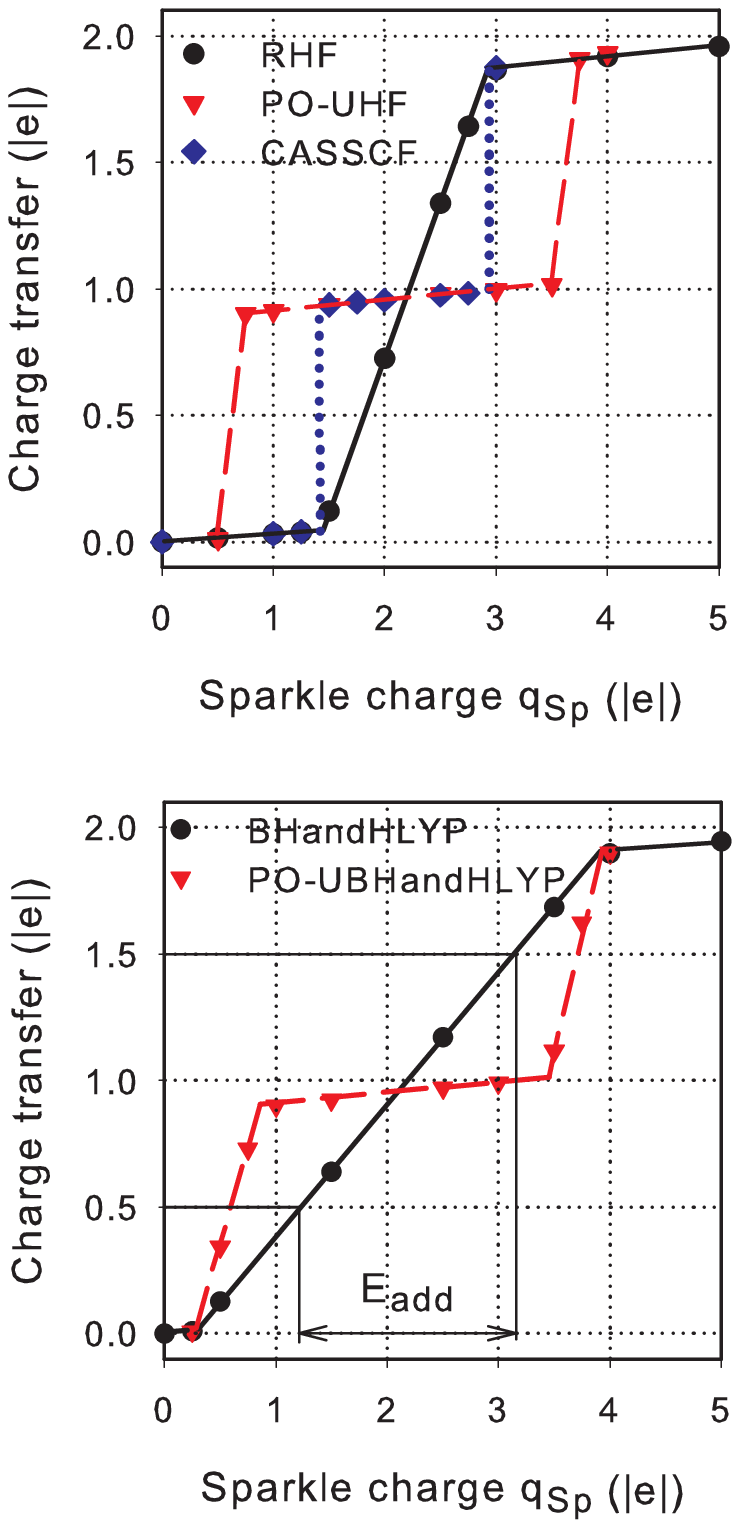}
      \caption[cap.Pt6]{\label{fig.PHP} (Color online) Intramolecular charge transfer between the donor and acceptor moieties in TCNQ-PHP-TTF induced by sparkle charges as obtained by a) {\it ab initio} wavefunction and b) hybrid DFT with a 6-31G** basis set. In a) restricted HF (circles), PO-unrestricted HF (triangles), and CASSCF (diamonds, with the HOMO and LUMO included in the active space) results are compared. In b) restricted (circles) and PO-unrestricted BHandHLYP (triangles) data are compared. The addition energies E$_{add}$ defined in Refs.~\cite{first} and~\cite{second} are also illustrated. In both a) and b), the BS-unrestricted initial guess leads to the closed-shell restricted solution.}
\end{figure}

Spin-unrestricted or spin-polarized calculations have been performed in the past for a wide variety of open-shell systems in quantum chemistry. Within wavefunction theory, it is well known for homolytic bond dissociation that although a multireference ansatz is the correct and general approach to the problem, a single-determinant spin-unrestricted Hartree-Fock (UHF) approach can provide a fair description, unlike RHF which produces qualitatively wrong results~\cite{SO96}. A spin-polarized solution allows initially paired electrons to sit in different spatial orbitals upon dissociation; however, this additional degree of freedom comes at the cost of a breaking in the symmetry of the many-electron wavefunction resulting in spin contamination, which is sometimes referred to as the symmetry dilemma in the literature~\cite{PSB95}. Spin-polarized (both pure and hybrid) DFT schemes are also known to describe dissociation problems quite satisfactorily~\cite{POLO02,GRAF04}. Their performance in describing open-shell molecular systems, such as singlet biradicals or dissociating molecules, classified as type-I and type-II systems, respectively, have also been explored~\cite{cremer1,cremer,GRAF02}.

It is therefore of interest to investigate whether single-determinant spin-unrestricted DFT and HF methods are able to deliver a step-like one-electron jump between weakly coupled donor and acceptor moieties upon charging by an external field, which is the crucial feature absent in closed-shell descriptions. For this purpose, we focus again on the two molecules introduced in the previous Section.  

It is well known with both UHF or spin-polarized DFT that, when the starting wavefunction or charge density in a self-consistent field (SCF) calculation is built in a conventional manner, the converged solution is always identical to the closed-shell result, even for cases where this solution is physically unstable. This is because the original symmetry of the initial guess cannot be broken by the self-consistency cycles, implying that if the spatial orbitals are initially the same for both spins, this will be also the case at the end. Thus, the spin-symmetry of the initial guess has to be broken in some way in order to reach the energetically lowered and correct open-shell solution for systems where such a solution is likely to exist.

Two ways for achieving this are commonly used~\cite{cremer}: (i) occupied and vacant orbitals are permuted for one spin only (permuted orbital scheme PO) thus creating directly an open-shell initial guess (this is implemented in Gaussian with the keyword guess=alter); or (ii) the HOMO and LUMO of the closed-shell initial guess are mixed (broken symmetry scheme BS), thereby breaking its spatial symmetry and introducing in the new guess some two-configurational character (this is implemented in Gaussian with the keyword guess=mix). For both schemes, some degree of spin contamination due to the mixture of singlet and triplet states before the SCF cycle is expected to be also found in the converged open-shell result. Nevertheless, the PO initial guess can offer a possibility to treat type-I systems at a single-determinant level while the BS initial guess can lead to reasonable results for type-II systems~\cite{cremer}.

Since the closed-shell HOMO and LUMO of the D-A system in our calculations correspond mainly to the HOMO of the donor and the LUMO of the acceptor, respectively, permuting or mixing them enforces charge transfer between these moieties in the initial guess and thus favor the desired solution. We stress that with both methods the initial guess engineering does not lead to an open-shell solution different from the closed-shell one in the absence of an external field (Figs.~\ref{fig.PP} and~\ref{fig.PHP}). For both molecules, DFT calculations with LDA or B3LYP functionals collapse into the closed-shell solutions whatever the initial guess. We thus focus hereafter on results obtained with Hartree Fock and hybrid DFT using the BHandHLYP functional.

For the molecule with partial overlap between donor and acceptor orbitals (TCNQ-PP-TTF in Fig.~\ref{fig.mo}), we find that both approaches to define the initial guess in open-shell systems (PO and BS) allow us to recover distinct steps in the electron transfer with UHF (Fig.~\ref{fig.PP}a) and spin-polarized hybrid DFT (Fig.~\ref{fig.PP}b). There are, however, significant differences in the stability range of the UHF results obtained with the two approaches compared to CASSCF, providing the correct benchmark solution with the lowest total energy (inset of Fig.~\ref{fig.PP}a). Some lowering of the CASSCF energies by a constant amount for 0 and 2 electron transfer (regions I and III in Fig.~\ref{fig.model}) with respect to RHF is due to the stabilization arising from closed-shell dynamic electron correlation. More significant is the CASSCF energy lowering in the region II, where the RHF solution is inadequate; the quality of the UHF solutions is determined by the proximity of their energies to the CASSCF energy curve. As can be seen from Fig. 8a and its inset, PO-UHF tends to deviate from CASSCF by inducing the electron transfer too early, which can be intuitively understood by the fact that, within this approach, the initial guess is already rather close to the solution for the full-electron transfer configuration. The BS-UHF method, with the initial guess configuration quite close to RHF, has the opposite problem so that the electron jump occurs for a larger field compared to CASSCF. 

For both approaches, the UHF solutions are reasonable approximations to CASSCF only in the narrow range of field where their total energies coincide (see inset of Fig.~\ref{fig.PP}a) though they are even there spin contaminated with $\braket{S^2} \sim$ 1.3. The open-shell solutions from spin-polarized DFT with the BHandHLYP functional for the same molecule (Fig.~\ref{fig.PP}b) do not appear to suffer from the same deficiencies. In this case, both PO and BS converge to similar solutions that are also quite close to the CASSCF  benchmark in Fig.~\ref{fig.PP}a. 

For the non-conjugated TCNQ-PHP-TTF molecule (Fig.~\ref{fig.mo}), the BS-UHF solution coincides with RHF (Fig.~\ref{fig.PHP}a). Since there is here no overlap between the HOMO and LUMO levels which are localized on the rather distant D and A moieties, the assumption of BS-UHF that only small corrections to the RHF initial guess would be sufficient to direct the open-shell solution simply fails. On the other hand, PO-UHF (Fig.~\ref{fig.PHP}a) and spin-polarized DFT with a BHandHLYP functional (Fig.~\ref{fig.PHP}b) do recover a step-wise electron transfer. These solutions are not spin pure and point to singlet-triplet mixtures with $\braket{S^2}$ of about 1.3 and 1, for UHF and spin-polarized DFT, respectively. For a quantitative assessment of the usefulness of our results, we need to compare them to the correct pure singlet state described by CASSCF for the whole range of external field. Fig.~\ref{fig.PHP}a shows that the PO-UHF step occurs at a field threshold much lower than for CASSCF (in close similarity with Fig.~\ref{fig.PP}a), whereas it occurs at the onset of partial electron transfer in RHF, as explained in Section~\ref{sec:twomol}. The behavior of the spin-polarized DFT results (Fig.~\ref{fig.PP}b) appears to be rather similar to PO-UHF (Fig.~\ref{fig.PP}a) for TCNQ-PHP-TTF, which puts it at odds with the CASSCF reference.

We emphasize again that our results obtained with LDA or BLYP as well as with the quite commonly used hybrid functional B3LYP (which we do not show here), exhibit artificial continuous charging even when allowing for spin-polarization. Whatever the choice of the initial guess, with these parameterizations of the XC functional, the open-shell ansatz collapses in general wrongly into the closed-shell solution for the TCNQ-PP-TTF molecule. In the absence of any orbital overlap, which is the case e.g. for TCNQ-PHP-TTF, we encounter spurious behavior in the SCF cycle and convergence problems. It is only with BHandHLYP (50\% of HF exchange) that we can observe the physically correct transfer of integer amounts of electrons.

As a summary for this subsection, we point out that spin-polarized hybrid DFT calculations both with PO and BS initial guesses appear to be a viable solution for describing stepwise electron transfer with quite good threshold values for the external field, at least for the TCNQ-PP-TTF molecule with a strong conjugation between the D and A units. For TCNQ-PHP-TTF characterized by a smaller coupling, the transfer steps are correctly described at a qualitative level whereas the threshold field values deviate quite substantially from the correct reference provided by CASSCF. The performance of UHF are quantitatively not very convincing for both molecules although steps are also observed in the curves with the BS and PO approaches. This failure of a Hartree Fock ansatz is a further evidence that the problems encountered with single-determinant closed-shell methods for the description of electron transfer in the weak coupling regime are not necessarily intimately linked to the self-interaction issue in DFT.

\end{subsection}

\begin{subsection}{The role of the HOMO-LUMO gap in electron transfer processes}
\begin{figure}
      \includegraphics[width=0.90\linewidth,angle=0]{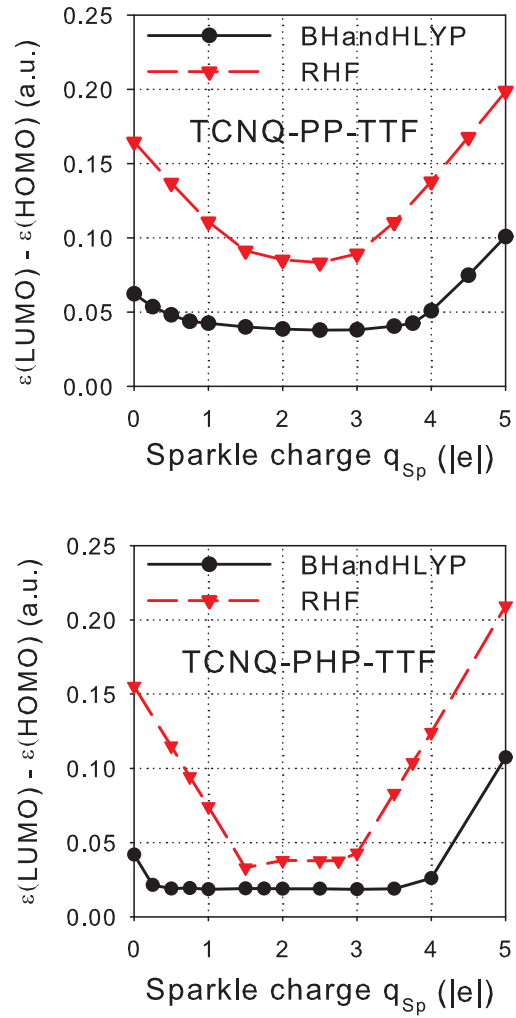}
      \caption[cap.Pt6]{\label{fig.gap} (Color online) HOMO-LUMO gap as a function of the sparkle charge for a) TCNQ-PP-TTF and b) TCNQ-PHP-TTF at the closed-shell Hartree-Fock (open symbols) and hybrid DFT BHandHLYP (closed symbols) levels. }
\end{figure}

The HOMO-LUMO gap in a closed-shell Hartree-Fock description is an important parameter that fully determines the thresholds or onsets for electron transfer processes. Our analytical model introduced in Sec.~\ref{sec:twomol} predicts, in the absence of overlap between the frontier orbitals, that the Coulomb integral is equal to the energy difference between the two levels at the threshold value of the external field (see Eq.~\ref{eqn.cond}). We now address here whether our model predictions are reflected in the evolution of the HOMO-LUMO gap with the external field, as evaluated from first-principle calculations both in the framework of HF and DFT with hybrid XC functionals. We show in Figs.~\ref{fig.gap}a and b for the TCNQ-PP-TTF and TCNQ-PHP-TTF molecules, respectively, the evolution of the HOMO-LUMO gap provided by closed-shell RHF calculations and hybrid DFT with a BHandHLYP functional as a function of the external field. The gap decreases with a growing external field up to a certain threshold at which electron transfer is initiated and then remains constant until the two electrons have been exchanged between the frontier orbitals, in full consistency with the analytical model. In close similarity with the differences observed between the charging curves of the two molecules in Figs.~\ref{fig.PP} and~\ref{fig.PHP}, the shape of the curves in Fig.~\ref{fig.gap}a for TCNQ-PP-TTF is found to be somewhat smoothened due to orbital overlap when compared to the strictly linear behaviour obtained for TCNQ-PHP-TTF in Fig.~\ref{fig.gap}b. 

For the threshold gap $\Delta\epsilon_c$, which we define as the size of the HOMO-LUMO gap at the onset of orbital mixing, we obtain values of 2.26 and 1.03 eV from HF and 1.03 and 0.52 eV from hybrid DFT for TCNQPP-TTF and TCNQ-PHP-TTF, respectively, from Fig.~\ref{fig.gap}. In  Fig.~\ref{fig.distance} we showed that the asymptotic limit of this threshold evolves as $1/d$ for large $d$ within the framework of HF. Making use of this linear relationship we can formally calculate from these critical gaps effective electron transfer distances as 
\begin{eqnarray}
\label{eqn.deff}
d_{eff}(HF)=\frac{1}{\Delta\epsilon_c(HF)}, 
\end{eqnarray} 
where atomic units have been assumed and related conversion factors have been omitted. In this way we derive $d_{eff}$ to be 6.4 \AA\ for TCNQ-PPTTF and 13.9 \AA\ for TCNQ-PHP-TTF, which roughly corresponds to the separation between the centers of the TTF and TCNQ moieties (5.9 \AA\ for the PP and 14.2 \AA\ for the PHP bridge, respectively). 

Figs.~\ref{fig.gap}a and b also show that, when moving from a pure HF description to DFT with a BHandHLYP XC functional containing 50\% of HF exchange, $\Delta\epsilon_c$ decreases by the same factor, i.e., by one half for TCNQ-PHP-TTF and a close value for TCNQ-PP-TTF. This is not obvious to explain since the meaning of the electronic eigenenergies of MOs is different within HF and DFT. In pure DFT (with semilocal exchange) these eigenvalues correspond to chemical potentials and charge transfer should only take place when the energies of the frontier orbitals are actually crossing due to the shift induced by the external field. In HF, on the other hand, charge transfer sets in when they are still separated by a threshold gap compensating the Coulomb integral between the two orbitals as we have shown in Section~\ref{sec:twomol}A. 

The virtually linear relation between the threshold MO gap and the percentage of HF/DFT exchange obtained from Fig.~\ref{fig.gap} now allows us to generalize Eq.~\ref{eqn.deff} derived so far within HF by introducing the fraction of HF exchange in the XC functional of hybrid DFT $f_{HF-X}$ as a variable. We then obtain:
\begin{eqnarray}
\label{eqn.hybrid}
\Delta\epsilon_c \sim \frac{f_{HF-X}}{d_{eff}},
\end{eqnarray}
which is valid for HF ($f_{HF-X}$=1), DFT with semilocal exchange ($f_{HF-X}$=0) and by design for hybrid DFT, where $f_{HF-X}$ is defined by the choice of XC functional.

For all discussed methods, the threshold external potential for electron transfer $\Delta\epsilon_c$ is physically defined by: i) the energy gap between the relevant donor and acceptor orbitals of the system without the perturbation of the external field; and ii) the slope of the MO energies as a function of the field which is purely electrostatic in its origin~\cite{rect1,rect2}. Since the HOMO-LUMO gap (or bandgap in solids) is severely underestimated by DFT with local or semi-local XC functionals but, on the other hand, overestimated by HF, hybrid DFT schemes provide in many cases a solution which better agrees with experimental data~\cite{paier}.

\end{subsection}

\begin{subsection}{Practical schemes for calculating addition energies from closed-shell DFT calculations}

We have demonstrated above that the qualitatively correct step-like behavior of charging curves can be obtained from the open-shell (spin-polarized) solutions of DFT with hybrid functionals, though the quantitative values for the onsets still differ from the CASSCF reference. There is, however, a different way to approach the problem within DFT, where it becomes possible to derive quantitatively realistic values for CB addition energies from closed-shell calculations although the charging curves are qualitatively wrong. 

In order to explain this apparent contradiction it has to be emphasized that the HOMO and LUMO in a single-particle KS scheme does not match in general the total-energy difference between the ground state and lowest charged states when the size of the HOMO-LUMO gap is finite~\cite{kohn}. This has been recently exploited for realistic calculations of  E$_{add}$ with standard DFT techniques in three different ways: i) for metal particles of finite size, a modified KS gap has been suggested, where the energetic difference of the LUMO for charged and uncharged clusters has been directly taken into account~\cite{capelle}; ii) for the description of the gap at C$_{60}$-metal interfaces, the charging energy has been obtained by using a constrained DFT formalism~\cite{louie}, where the occupation of hand-picked orbitals can be defined as a constraint in the input~\cite{wu}; iii) within a NEGF-DFT framework, E$_{add}$ has been defined via threshold values of an external gate voltage V$_{gate}$ determined via a midpoint integration rule from induced charge transfer between small molecules (H$_2$ and benzene) and lithium wires~\cite{first}; this method has been also extended to aluminum surfaces and was shown to correctly describe screening effects~\cite{second}. According to the midpoint integration rule in Refs.~\cite{first,second}, the onsets can be determined from the voltages required to move 0.5 and 1.5 electrons, respectively. As a matter of fact, Figs.~\ref{fig.PP} and~\ref{fig.PHP} show that the onsets determined from the closed-shell DFT solutions match almost exactly the CASSCF results. 

\end{subsection}

\end{section}

\begin{section}{Summary}\label{sec:summary}

In conclusion, we have presented a configuration interaction description of electron transfer between weakly coupled organic molecules. The use of a multideterminant wavefunction approach points to the key role of many-body effects in such a scenario. Our approach yields the physically correct step-like features while a closed-shell ansatz, either in the DFT or RHF framework, introduces severe limitations that we have explained via an analytical derivation. First-principle calculations in the HF and DFT framework corroborate our predictions for the deficiencies of closed-shell solutions. In a proper open-shell treatment, step-like jumps that are the hallmark of charge quantization can be recovered with UHF and spin-polarized DFT with a hybrid exchange. Furthermore, by relating the onset of charge transfer to the zero bias HOMO-LUMO gap, we discussed its origin and meaning in the context of both HF and DFT. Our work has also been connected to recently proposed practical schemes for calculating the addition energies in electron transport experiments in the CB regime from closed-shell DFT calculations.

\end{section}

\begin{section}{Acknowledgments}

This research has been supported by the European Commission with the projects SINGLE (FP7/2007-2013, no. 213609), MODECOM (NMP3-CT-2006-016434) and MINOTOR (FP7-NMP-228424), the Interuniversity Attraction Pole IAP 6/27 Program of the Belgian Federal Government, and the Belgian National Fund for Scientific Research (FNRS). J.C. in particular is funded by the FNRS. R.S. is currently supported by the Austrian Science Fund FWF, project Nr. P20267.

\end{section}


\bibliographystyle{apsrev}

\end{document}